\documentclass[aps,preprint,showpacs, showkeys]{revtex4}%
\usepackage{amsfonts}
\usepackage{amsmath}
\usepackage{amssymb}
\usepackage{graphicx}%
\providecommand{\U}[1]{\protect\rule{.1in}{.1in}}

\begin{document}
\title[ ]{Exact Metric Operators as the Ground State functions of the Hermitian
Conjugates of a Class of Quasi-Hermitian Hamiltonians }
\author{Abouzeid. M. Shalaby\thanks{E-mail:amshalab@ mans.edu.eg}}
\affiliation{Physics Department, Faculty of Science, Mansoura University, Egypt.}

\begin{abstract}
We generalized a class of non-Hermitian Hamiltonians which introduced
previously by us in such a way in which every member in the class is non-\textit{PT}-symmetric.
For every member of the class, the ground state is a constant with zero energy
eigen value. Instead of using an infinite set of coupled operator equations to
calculate the metric operator we used a simple realization to obtain the class
of closed form metric operators corresponding to the class of non-Hermitian
and non-\textit{PT}-symmetric Hamiltonians introduced. The trick is that, if
$\psi$ is an eigen function of $H$, then $\phi=\eta\psi$ is an eigen function
of $H^{\dagger}$ with the same eigen value. Thus, knowing any pair $(\psi
,\phi)$ one can deduce the form of the exact metric operator. We note that,
the class of Hamiltonians generalized in this work has the form of that of
imaginary magnetic field which can be absorbed by the quasi-gauge
transformations represented by metric operators. Accordingly, it is expected
that the $Q$ operators will disappear for the whole members in the class in
the path integral formulation. However, the detailed analysis of this issue
will appear in another work.

\end{abstract}
\maketitle

The recent growing of researchers interest in the quasi-Hermitian Hamiltonians
leaded \ to the well establishment of such kind of theories
\cite{PT1,PT2,PT3,PT4,PT5,PT6,PT7,bender1,spect,spect1,bender,abo}.  Although
there exists a Hermitian Hamiltonian which is equivalent to each non-Hermitian
Hamiltonian with real spectrum, one can not conclude that such kind of
research is redundant. In fact, there is a plenty of benefits in studying the
theory in its non-Hermitian representation because in many cases it is simpler
than the Hermitian representation. For instance, the non-Hermitian $\phi^{4}$
scalar field theory is a plausible candidate to play the role of the scalar
sector in the standard model for particle interactions \cite{abo1} while its
Hermitian equivalent Hamiltonian is not known till now. Also, we have showed
that the Hermitian $\phi^{6}$ field theory with coupling of negative two mass
dimension, can be converted to an equivalent non-Hermitian representation in
which the coupling has a negative one mass dimension \cite{abo2}  which is a very important result
toward the solution of the unification problem. Besides, the equivalent
non-Hermitian representation is a $\phi^{4}$-like theory for which the Feynman
diagram calculations go more simpler than in the Hermitian $\phi^{6}$ field
theory. According to these benefits, one can say that at least, the
quasi-Hermitian representation can be used as a calculational algorithm in
both quantum mechanics and quantum field theories. Also, the subject as a
whole is needed to discover if a non-Hermitian theory is physically acceptable
or not thorough the investigation of the possible existence of a positive
definite metric operator.

The metric operator has a vital role in the study of quasi-Hermitian theories
\cite{spect,spect1}. However, there exist rare cases for which the metric
operator has been obtained in an exact manner \cite{abo3,jones}. Even in this
case, one has to solve a set of an infinite number coupled operator equations
and one needs to be lucky enough to have a truncation at a small order in the
perturbation expansion used. In this letter, we introduce a some how different
way to obtain the closed form metric operator for a generalized class of
non-Hermitian Hamiltonians. We mean by generalized that, in this work we
generalize the form of a new class of non-Hermitian Hamiltonians introduced
previously by us. In the generalized form, every member is non-Hermitian as
well as non-PT-symmetric rather than the previous new class in Ref.\cite{abo4} for which
half of the members are PT-symmetric .

To start, consider a Hamiltonian $H$ which has a set of eigen functions
$\left\{  \psi_{n}\right\}  $. The Hamiltonian $H$ is said to be $\eta$-Pseudo
Hermitian if $H^{\dagger}=\eta H\eta^{-1}$, where $\eta$ is a Hermitian
invertible linear operator. In fact, the $\eta$ operator is not unique and if
the set ${\varepsilon}\left(  H\right)  $, the set of all $\eta$'s, includes
some $\eta$ such that the inner product defined by $\langle\psi|\eta
|\psi\rangle=\langle\langle\psi|\psi\rangle\rangle_{\eta}$ is positive
definite, then $H$ is Hermitian in the Hilbert space endowed with the
$\langle\langle\psi|\psi\rangle\rangle_{\eta}$ inner product and thus the
spectrum is real \cite{PT6,spect,spect1}.

To shed light on previous calculational procedures of the the positive
definite operator $\eta_{+}$, consider the Hamiltonian $H$ such that
$H|\psi_{n}\rangle=E_{n}|\psi_{n}\rangle$ \ then $H^{^{\dagger}}|\phi
_{n}\rangle=E_{n}|\phi_{n}\rangle$ and
\bigskip$\langle\phi_{n}|\psi_{m}\rangle=\delta_{nm}$,$%
{\displaystyle\sum\limits_{n}}
|\psi_{n}\rangle\langle\phi_{n}|=1.$ \newline The set $\left\{  |\psi
_{n}\rangle\text{, }\langle\phi_{n}|\right\}  $ form a biorthonormal system
and $H=%
{\displaystyle\sum\limits_{n}}
E_{n}|\psi_{n}\rangle\langle\phi_{n}|$, $H^{\dagger}=%
{\displaystyle\sum\limits_{n}}
E_{n}|\phi_{n}\rangle\langle\psi_{n}|$. In Ref.\cite{spect1}, it is deduced
that the operator $\eta_{+}$ can be represented as
\begin{equation}
\eta_{+}=%
{\displaystyle\sum\limits_{n}}
|\phi_{n}\rangle\langle\phi_{n}|,
\end{equation}
which is a positive-definite operator, satisfies the condition $H^{\dagger
}=\eta_{+}H\eta_{+}^{-1}$ and is a difficult infinite sum too.

In Bender 's regime, the inner product is defined through the introduction of
a $C$ operator, which is represented in the coordinate space as the sum%
\begin{equation}
C(x,y)=%
{\displaystyle\sum\limits_{n}}
\phi_{n}(x)\phi_{n}(y)\text{,} \label{cop}%
\end{equation}
where $\left\{  \phi_{n}(x)\right\}  $ are the coordinate-space eigen
functions of the Hamiltonian \cite{bender}. Because it is impossible to use
the sum in Eq.(\ref{cop}) for the calculation of the $C$ operator, Bender
\textit{et.al \ }introduced a powerful method for calculating $C$ by seeking
an operator representation of $C$ in the from $C=\exp(-Q(x,p)P$, where $P$ is
the parity operator. \ The calculation procedure is based on the observations
that%
\begin{equation}
\lbrack C,PT]=0\text{, }\ C^{2}=1\ \ \text{and }[C,H]=0.
\end{equation}
Accordingly, if we assume the Hamiltonian to take the form%
\begin{equation}
H=H_{0}+\epsilon H_{1},
\end{equation}
 and $Q(x,p)$ has the perturbative expansion%
\begin{equation}
Q(x,p)=\epsilon Q_{1}(x,p)+\epsilon^{3}Q_{3}(x,p)+......
\end{equation}
then the first three operator equations are given by%
\begin{align}
\lbrack H_{0},Q_{1}]  &  =-2H_{1}\text{,}\\
\lbrack H_{0},Q_{3}]  &  =-\frac{1}{6}[Q_{1},[Q_{1},H_{1}]]\text{,}\\
\lbrack H_{0},Q_{5}]  &  =\frac{1}{360}[Q_{1},[Q_{1},[Q_{1},H_{1}]]]-\frac
{1}{6}[Q_{1},[Q_{3},H_{1}]]-\frac{1}{6}[Q_{3},[Q_{1},H_{1}]].
\end{align}
 Using such a representation they were able to calculate $C$ up to
$\epsilon^{7}$ for the non-Hermitian model
\begin{equation}
H=\frac{1}{2}p^{2}+\frac{1}{2}\mu^{2}x^{2}+i\epsilon x^{3},
\end{equation}
and up to $g^{1}$ for the quantum field version of the form%
\begin{equation}
H=\frac{1}{2}\left(  \partial\phi\right)  ^{2}+\frac{1}{2}m^{2}\phi^{2}%
+ig\phi^{3}. \label{field}%
\end{equation}
For the $\left(  -g\phi^{4}\right)  $, we need a non-perturbative tool ( WKB,
for instance).

So, for most of the theories, one need to duplicate the effort to obtain a
well established formulation of a non-Hermitian theory with real spectra. At
one hand, we calculate the wave functions up to some order and on the other
hand one has to calculate the metric operator or the $C$ operator up to the
same order. Here we suggest another route for the calculation of the positive definite metric operator. In fact, a simple realization is that if we have a
non-Hermitian Hamiltonian such that there exists a linear invertible operator
$\eta$ satisfying
\begin{equation}
\eta H\eta^{-1}=H^{\dagger}\text{,}%
\end{equation}
and if $\psi_{n}$ is an eigen function of $H$ with eigen value $E_{n}$ then%
\begin{align}
H^{\dagger}\left(  \eta\psi\right)   &  =\eta H\eta^{-1}\left(  \eta
\psi\right) \\
&  =\eta H\psi=E_{n}\left(  \eta\psi\right)  ,
\end{align}
\textit{i.e }$\eta\psi$ is an eigen function of $H^{\dagger}$ with the same
eigen value. Accordingly,  by solving the Shr\"{o}dinger
equations for both $H$ and $H^{\dag}$, and the through the relation
\begin{equation}
\phi=\eta\psi,
\end{equation}
where $\phi$ is the eigen function of $H^{\dag}$, 
 one can deduce the form of $\eta$.
To clarify the point, consider the non-Hermitian Hamiltonian of the form%

\begin{equation}
-\frac{d^{2}\psi}{dx^{2}}+\left(  2ixp\right)  \psi=E\psi,
\end{equation}
where $p$ is the momentum operator. Clearly, the constant $c$ is an eigen
function of zero eigen value. Now consider the the corresponding
Shr\"{o}dinger equation for $H^{\dag}$%

\begin{equation}
-\frac{d^{2}\phi}{dx^{2}}-\left(  2ipx\right)  \phi=E\phi,\nonumber
\end{equation}
which has the eigen function $c\exp\left(  -x^{2}\right)  $. Accordingly,
$\eta=\exp\left(  -x^{2}\right)  $. To check that this really the metric
operator consider%
\begin{equation}
\exp\left(  -x^{2}\right)  H\exp\left(  x^{2}\right)  =p^{2}-2ixp-2=H^{\dag}.
\end{equation}
Also, consider the relation
\begin{equation}
\exp\left(  \frac{-x^{2}}{2}\right)  H\exp\left(  \frac{x^{2}}{2}\right)
=p^{2}+x^{2}-1,
\end{equation}
and thus the Hermitian Hamiltonian $h=p^{2}+x^{2}-1$ is equivalent to the
non-Hermitian Hamiltonian $p^{2}+2ixp$ which shows that $\eta=\exp\left(
-x^{2}\right)  $ passed all the tests as a positive definite metric operator.

Now consider the non-Hermitian and non-PT-Symmetric Hamiltonian of the form%
\begin{equation}
H=\frac{1}{2}p^{2}+i\left(  gx^{2}+\omega x+d\right)  p,
\end{equation}
Clearly, the constant $c$ represents the ground state function of $H$.
\ Moreover, the function $\phi=c\exp(-\frac{2gx^{3}}{3}-\omega x^{2}-2dx)$
represents the ground state function for $H^{\dag}$ and the corresponding
Hermitian Hamiltonian $h$ takes the form
\begin{equation}
h=\frac{1}{2}p^{2}+\left(  gd+\frac{1}{2}\omega^{2}\right)  x^{2}+\frac{1}%
{2}g^{2}x^{4}+gmx^{3}+\left(  \omega d-\allowbreak g\right)  x-\frac{1}%
{2}\omega+\frac{1}{2}d^{2}\text{.}%
\end{equation}
To assure our result, let us consider one more example for the Hamiltonian%
\begin{equation}
H=\frac{1}{2}p^{2}+i\left(  fx^{3}+gx^{2}+\omega x+d\right)  p,
\end{equation}
with the ground state function $c$ and the the ground state
\begin{equation}
\phi=c\exp(-\frac{1}{2}fx^{4}-\frac{2}{3}gx^{3}-\omega x^{2}-2dx)
\end{equation}
for $H^{\dag}$ and
\begin{align}
h  &  =\exp(\frac{-\frac{1}{2}fx^{4}-\frac{2}{3}gx^{3}-\omega x^{2}-2dx}%
{2})H\exp(-\frac{-\frac{1}{2}fx^{4}-\frac{2}{3}gx^{3}-\omega x^{2}-2dx}{2})\\
&  =\frac{1}{2}p^{2}+\left(  \frac{1}{2}\omega^{2}+gd-\frac{3}{2}f\right)
x^{2}+\left(  fd+g\omega\allowbreak\right)  x^{3}+\left(  \frac{1}{2}%
g^{2}+f\omega\right)  x^{4}+fgx^{5}\\
&  +\frac{1}{2}f^{2}x^{6}+\left(  \omega d-g\right)  x-\frac{1}{2}\omega
+\frac{1}{2}d^{2}.
\end{align}
 In general, a non-Hermitian Hamiltonian of the form
\begin{equation}
H_{M}=\frac{1}{2}p^{2}+i\left(  \sum_{n=0}^{M}a_{\pm n}x^{\pm n}\right)  p,
\label{class}%
\end{equation}
where $M$ is a positive integer, has a real spectrum and the exact metric
operator for this Hamiltonian is given by

\bigskip%
\begin{equation}
\eta=\exp\left(  -\sum_{n=0}^{M}\frac{2a_{\pm n}x^{\pm n+1}}{\left(  \pm
n+1\right)  }\right)  ,
\end{equation}
and the corresponding Hermitian Hamiltonian takes the form;%
\begin{equation}
h_{M}=\exp\left(  -\sum_{n=0}^{M}\frac{a_{\pm n}x^{\pm n+1}}{\left(  \pm
n+1\right)  }\right)  H_{M}\exp\left(  \sum_{n=0}^{M}\frac{a_{\pm n}x^{\pm
n+1}}{\left(  \pm n+1\right)  }\right)  .
\end{equation}
For example, when $M=6$ and we take positive $n$ only in $H_{M}$ we get the
equivalent Hermitian Hamiltonian of the form%
\begin{align}
h_{6}  &  =\frac{1}{2}a_{6}^{2}x^{12}+a_{5}x^{11}a_{6}+\left(  \frac{1}%
{2}a_{5}^{2}+a_{4}a_{6}\right)  x^{10}+\allowbreak\left(  a_{4}a_{5}%
+a_{3}a_{6}\right)  x^{9}+\left(  a_{2}a_{6}+\frac{1}{2}a_{4}^{2}+a_{3}%
a_{5}\right)  x^{8}\\
&  +\allowbreak\left(  a_{3}a_{4}+a_{2}a_{5}+a_{1}a_{6}\right)  x^{7}+\left(
a_{0}a_{6}+a_{2}a_{4}+\frac{1}{2}a_{3}^{2}+a_{1}a_{5}\right)  \allowbreak
x^{6}\\
&  +\left(  -3a_{6}+a_{0}a_{5}+a_{1}a_{4}+a_{2}a_{3}\right)  x^{5}+\left(
\frac{1}{2}a_{2}^{2}-\frac{5}{2}a_{5}+a_{1}a_{3}+a_{0}a_{4}\right)
\allowbreak x^{4}\\
&  +\left(  a_{1}a_{2}+a_{0}a_{3}-2a_{4}\right)  x^{3}+\left(  -\frac{3}%
{2}a_{3}+a_{0}a_{2}+\frac{1}{2}a_{1}^{2}\right)  \allowbreak x^{2}+\left(
-a_{2}+a_{0}a_{1}\right)  x\\
&  -\frac{1}{2}a_{1}+\frac{1}{2}a_{0}^{2}+\frac{1}{2}p^{2}.
\end{align}
Note that, since the exact metric operator for the above mentioned
Hamiltonians is a function of $x$ only then the more general Hamiltonian of
the form%
\begin{equation}
H_{M}=\frac{1}{2}p^{2}+i\left(  \sum_{n=0}^{M}a_{\pm n}x^{\pm n}\right)
p+f(x),
\end{equation}
where $f(x)$ is a polynomial in $x$ is quasi-Hermitian and have the metric
operator
\begin{equation}
\eta=\exp\left(  -\sum_{n=0}^{M}\frac{2a_{\pm n}x^{\pm n+1}}{\left(  \pm
n+1\right)  }\right)  ,
\end{equation}
and the equivalent Hermitian Hamiltonian is given by%
\begin{equation}
h_{M}=\exp\left(  -\sum_{n=0}^{M}\frac{a_{\pm n}x^{\pm n+1}}{\left(  \pm
n+1\right)  }\right)  H_{M}\exp\left(  \sum_{n=0}^{M}\frac{a_{\pm n}x^{\pm
n+1}}{\left(  \pm n+1\right)  }\right)  .
\end{equation}
Accordingly, knowing only one wave function out of the spectra of both $H$ and
$H^{\dagger\text{ }}$yields in an automatic way the closed form positive
definite metric operator provided that both $H$ and $H^{\dagger\text{ }}$have
the same real spectra.

The above formulations shows that, the non-Hermitian representations in
quantum mechanics is a powerful calculational tool for certain cases as it
turns out a Hermitian Hamiltonian which is a polynomial of order $2M$ \ in the
creation and annihilation operators into an equivalent non-Hermitian
Hamiltonian which is a polynomial of degree $M$ \ in $a$ and $a^{\dagger}$. As
we know, in perturbation calculations, the number of non-vanishing matrix
elements increases as the highest power in $a$ and $a^{\dagger}$ increases and
thus working in the non-Hermitian representation will lower the number of
non-vanishing matrix elements and thus simplifies the calculations.

For the generalized class in this work, the ground state function of the
non-Hermitian Hamiltonian is a constant for each member in the class with the
ground state energy is zero. Accordingly, the ground state function for the
corresponding Hermitian Hamiltonian can be taken as $\rho=\sqrt{\eta}%
=\exp\left(  -\sum_{n=0}^{M}\frac{2a_{\pm n}x^{\pm n+1}}{2\left(  \pm
n+1\right)  }\right)  $. Thus, for $M$ even and for positive $n$ values only,
the ground state function is not square integrable and thus does not represent
a true Physical state. On the other hand, for $M$ odd, the ground state is
square integrable.

What is amazing in this work that it relates Hermitian Hamiltonians of
polynomial form to a non-Hermitian Hamiltonians with velocity dependent
potentials. In fact, the non-Hermitian Hamiltonians in this work have the form
of a theory of imaginary magnetic field. Moreover, the metric operator
represents a quasi-quage transformation which adds a terminal term to the
action and thus it is expected that the $Q$ operator will disappear in the
path integral formulation. However, we postpone such kind of investigations to
another work.

To conclude, we generalized a form of a class of quasi-Hermitian theories in
such a way that each member in the class is non-PT-symmetric. We used a simple
method to obtain the cosed form metric operator for each member in the class.
In this method, since we realized that for each member the ground state is a
constant, we solved the corresponding Shr\"{o}dinger equation for $H^{\dagger
}$ with $E=0$, which then is nothing but the metric operator $\eta$. In fact,
the Shr\"{o}dinger equation for $H^{\dagger}$have a simple shape of the well
known Fokker-Plank equations for which one can get the exact solution for
$E=0$.

We assert that all the members of the class are real line theories and thus
makes the calculations simple. Also, the interaction terms has the form of a
particle in an imaginary magnetic field and the operator operator as well as
the operator $\rho=\sqrt{\eta}$ represent a quasi-gauge transformations which
means that one can work with any of the operator $H$, $H^{\dagger}$and $h$ in
the path integral formulations with the results stay the same. We will make
the explicit calculations regarding this issue and present it in another work.

The ground states for the class of Hermitian Hamiltonians $h$ are not all
physically acceptable. in fact, the ground state function for the
corresponding Hermitian Hamiltonian can be taken as $\rho=\sqrt{\eta}%
=\exp\left(  -\sum_{n=0}^{M}\frac{2a_{\pm n}x^{\pm n+1}}{2\left(  \pm
n+1\right)  }\right)  $. Thus, for $M$ even and for positive $n$ values only,
the ground state function are not square integrable and thus does not
represent a true Physical state. On the other hand, for $M$ odd, the ground
states are square integrable. Thus, it is not correct to consider each
non-Hermitian with real spectrum to be Physically acceptable.

\section*{Acknowledgment}

\label{ack} The author would like to thank Dr. S.A. Elwakil for his support
and kind help.


\end{document}